\documentclass[journal = jacsat, manuscript=article, layout=traditional]{achemso}
\setkeys{acs}{articletitle = true}

\usepackage{float}
\usepackage{natbib}
\usepackage{setspace}
\usepackage{xkeyval}

\author{Prince Gollapalli}
\author{Varalakshmi Jalligampala}
\author{Kishor Peddapuvvala}
\author{Prajeet Oza}
\author{Satyesh Kumar Yadav}
\email{satyesh@iitm.ac.in}
\affiliation[IITM]{Department of Metallurgical and Materials Engineering, Indian Institute of Technology (IIT) Madras, Chennai 600036, India.}
\alsoaffiliation{Center for Atomistic Modelling and Materials Design, Indian Institute of Technology (IIT) Madras, Chennai 600036, India.}

\title{3-D Ti/TiN Interface}
\SectionNumbersOn

\usepackage[utf8]{inputenc}
\usepackage{amsmath}
\usepackage{amsfonts}
\usepackage{amssymb}
\usepackage{graphicx}
\usepackage{gensymb}
\usepackage{multicol}
\usepackage{multirow}
\usepackage{hyperref}
\hypersetup{
	colorlinks=true, 
	linktoc=all,     
	linkcolor=magenta,  
	urlcolor=cyan,
	citecolor=blue
}

\begin{document}

\singlespacing

\maketitle
\section*{Keywords}
3-D metal/ceramic interface, mechanical properties, density functional theory

\section*{Abstract}

Interface by definition is two-dimensional (2-D) as it separates 2 phases with an abrupt change in structure and chemistry across the interface. The interface between a metal and its nitride is expected to be atomically sharp, as chemical gradation would require the creation of N vacancies in nitrides and N interstitials in metal. Contrary to this belief, using first-principles density functional theory (DFT), we establish that the chemically graded Ti/TiN interface is thermodynamically preferred over the sharp interface. DFT calculated N vacancy formation energy in TiN is 2.4 eV, and N interstitial in Ti is -3.8 eV. Thus, diffusion of N from TiN to Ti by the formation of N vacancy in TiN and N interstitial in Ti would reduce the internal energy of the Ti-TiN heterostructure. We show that diffusion of N is thermodynamically favorable till $\sim$23\% of N has diffused from TiN to Ti, resulting in an atomically chemically graded interface, which we refer to as a 3-D interface. Experiments' inability to identify a 3-D interface in Ti/TiN could be attributed to limitations in identifying chemical composition and structure with atomic-level resolution at interfaces. We define the sum of N vacancy formation energy and N interstitial formation energy as driving-force, which could be used as a convenient way to assess the possibility of forming a 3-D interface in metal/ceramic heterostructures. We also show gradual variation in lattice parameters and mechanical properties (like bulk modulus, shear modulus, Young’s modulus, and hardness) across the Ti/TiN interface. 3-D interfaces open a new way to control properties of metal/ceramic heterostructures, in line with the already established advantage of gradation at interfaces in micrometer length scale. For widely explored Ti/TiN multilayer nano-heterostructures, the possibility of forming 3-D interface could lead to enhanced wear and erosion resistance.

    
\section{Introduction}

\begin{figure}[tbp]
	\centering
	\includegraphics[width=\textwidth]{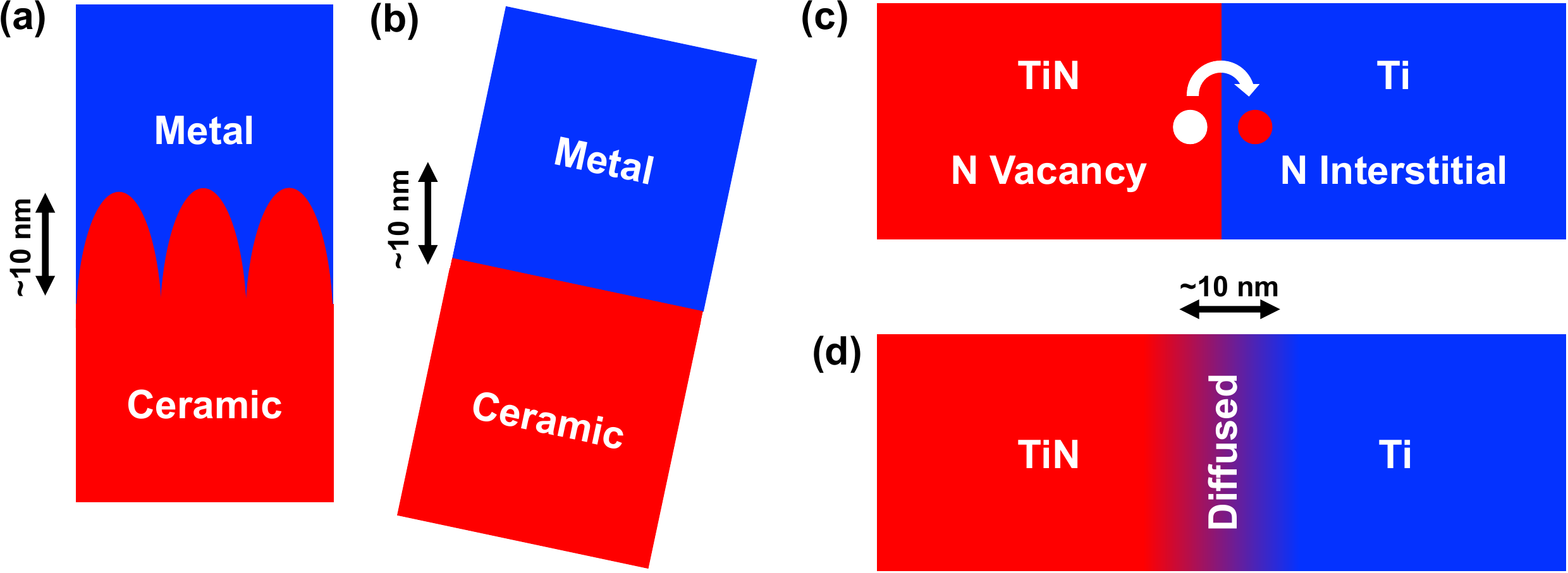}
	\caption{Schematic of the depth profile of metal/ceramic interfaces a) sharp and rough b) sharp and flat. (c) Schematic of sharp Ti/TiN interface, with the tendency of N to migrate from TiN to Ti. (d) Atomically chemically graded interface formed either due to migration of N from TiN to Ti or formation of compounds with a range of stoichiometry.}
	\label{figGradation}
\end{figure}

\paragraph{}
Interfaces play a critical role in the properties of nano-heterostructures because of the large volume fraction \cite{Beyerlein2015}. The structure of metal/ceramic interface, like misfit dislocation network, and crystallographic orientation, can significantly affect many properties \cite{Ruhle1989,Lackner2013,Finnis1996,Gutierrez2017} such as dislocation annihilation and nucleation, twinnability, recovery from radiation-induced defects, thermal, optical, and electrical conductivity \cite{Xiao2015}. These features can represent interface structure in flat and sharp interfaces. Atomically chemically graded interfaces present many more interface features like continually varying mechanical properties, chemical, and atomic structure. Chemical and microstructural gradation has been shown to improve wear and erosion properties \cite{Bonu2019,Swaminathan2010}, suppress crack formation in ceramic heterostructures, and improve resistance to contact deformation and damage \cite{Suresh2001,Zhang2021}. Similar effects could be expected due to atomically chemically graded interface for heterostructure.
\paragraph{}
There have been several efforts to create atomically graded interfaces in metal/ceramic heterostructures. All of them have focused on changing the ratio of metal and anions that form ceramic. For example, Raveh et al. intended to create a graded interface between Al and AlN by controlling the partial pressure of N \cite{Raveh1999}. Although it seems (based on AES depth profile) that they have achieved a graded interface, there are no reports of stable compounds with stoichiometry AlN$_x$(x$<$1). The reason for such observation could be a gradual change in the phase fraction of Al and AlN across the interface. Limitations of experimental tools used to measure the atomic fraction of materials across the interfaces invariably conclude that interfaces are graded. Fe/MgO interface is also shown to be graded despite MgO being well known to be a line compound; hence no intermixing of Fe in MgO should be expected  \cite{Li2017,Ljungcrantz1996}. The depth profile of the sharp and rough interface between metal and ceramic, as shown in Figure \ref{figGradation}(a), would be interpreted as chemically graded. Similarly, a sharp and flat interface, as shown in Figure \ref{figGradation}(b), could be construed as a chemically graded interface if the interface is tilted with respect to the line of sight. 
  
  \paragraph{}
  From a purely thermodynamic consideration, atomic-level chemical gradation can be achieved if 1) there are stable compounds with a range of stoichiometry or 2) mixing of anions or metal across the interface is thermodynamically favorable. There are only a handful of ceramics with a range of stoichiometry, for example, Ti-O system, Ti-C, Ti-N, and Fe-O, which we could find from equilibrium phase diagrams \cite{phaseDiagTi-O,phaseDiagTi-C,phaseDiagTi-N,phaseDiagFe-O}. Out of these, there has been extensive effort to form a chemically graded Ti/TiN interface by varying ratios of Ti and N during synthesis \cite{Watkins2016,Kupczyk1993}. Ti/TiN heterostructure finds a range of applications. Multilayer coating showed improved wear and corrosion protection on metals compared to bare TiN \cite{Subramanian2011,PatentPraxiar}. 
  
  \paragraph{}
  We would like to particularly mention Watkins {et al.} \cite{Watkins2016} work which showed a very smooth gradation in a mere distance of approximately 10 nm by controlling N partial pressure while growing the Ti/TiN heterostructure. This smooth gradation of N points to the possibility of a range of phases of Ti-N present at the interface. Although from the equilibrium phase diagram, only a few phases like solid solution of Ti and N with up to 20\% atomic solubility, Ti$_2$N, and TiN at room temperature are known \cite{WolffLodewijkBastinGiel1985}. Using DFT, we assess the relative stability of phases reported in the equilibrium phase diagram along with a few recently reported phases like Ti$_3$N$_2$, Ti$_4$N$_3$, and Ti$_6$N$_5$ \cite{Yu2015,Okamoto1993}. If these phases with increasing N concentration are indeed stable, they can form part of the graded interface.
  
  \paragraph{}
  Starting from a sharp interface, forming a chemically graded interface between metal and ceramics (oxide, nitride, or carbide) by diffusion of anions (O, N, or C) across the interface is not expected as it involves the formation of defects in both metal and ceramics. To our surprise, using first-principles density functional theory (DFT), we show that Ti favors assimilation of N atoms as interstitials within Ti over the formation of N vacancy in stoichiometric TiN. This assimilation of N atoms in Ti suggests that a graded interface is thermodynamically preferred over a sharp interface. The tendency of N to migrate from TiN to Ti in the initially sharp Ti/TiN is in line with the experimental observation by Susmita et al \cite{Chowdhury2021}. Figure \ref{figGradation} (c) schematically shows  N migration from TiN to Ti by forming N vacancies in TiN and N interstitials in Ti, starting from a sharp interface. If N migration continues until it becomes unfavorable, the resulting interface would be chemically graded, as shown in Figure \ref{figGradation} (d). We refer to this atomically chemically graded interface as a 3-D interface. 
    
  \paragraph{}
 Using first-principles DFT calculations, we show that a range of Ti-N systems is stable, explaining the formation of a chemically graded interface by controlling the growth parameters. Then we show that it is thermodynamically favorable for N to migrate from TiN to Ti, and approximately 23\% of N can diffuse from TiN to Ti. We also calculate misfit strains among various stable phases likely to be found in the 3-D interface and their mechanical properties.

  
\section{Methodology}
  	\subsection{First-principles Calculations}
  	
  \paragraph{}	
  	The first-principles electronic structure calculations were carried out within the density functional theory (DFT) \cite{Hohenberg1964,Kohn1965} as implemented in Vienna \textit{Ab initio} Simulation Package (VASP) code \cite{Kresse1996-1, Kresse1996-2}. The DFT calculations employed the Perdew, Burke, and Ernzerhof (PBE) \cite{Perdew1996} generalized gradient approximation (GGA) exchange-correlation functional and the projector-augmented wave (PAW) method \cite{Blochl1994}. For all calculations, a plane wave kinetic energy cutoff of 520 eV was used for the plane wave expansion of the wave functions to obtain highly accurate forces. The Brillouin zone sampling was performed using Monkhorst-Pack meshes \cite{Monkhorst1976} with a resolution of less than $2 \pi \times 0.03~\AA^{-1}$. All the structures were relaxed using conjugate gradient scheme, and the total energies were converged to less than 1 meV per atom. The forces acting on each atom are less than 0.01 eV/$\AA$. For Ti and N, 3p$^6$3d$^2$4s$^2$ and 2s$^2$2p$^3$ electrons were considered as valence electrons, respectively. All the crystal structures were drawn using VESTA \cite{Momma2011} software.

  	
\section{Results and Discussion}

  	  \subsection{Growth controlled formation of graded interface}

\begin{figure}[tp]
	\centering
	\includegraphics[width=10cm]{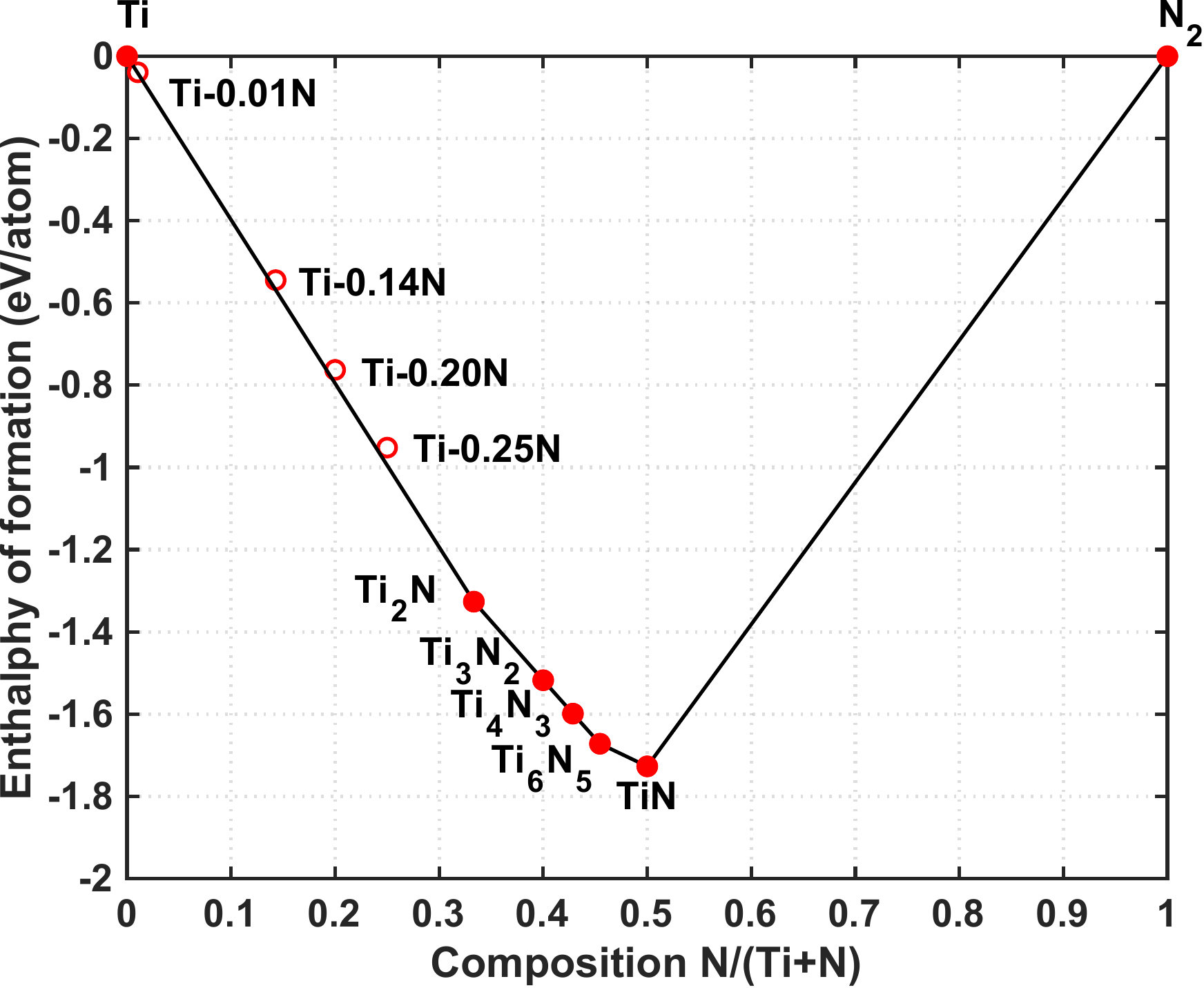}
	\caption{Convex hull diagram for the Ti-N system. Solid circles represent stable phases, and open circles represent phases above the hull.}
	\label{figHull}
\end{figure}

\paragraph{}
If graded interface forms by controlling N concentration, Ti-N phases with a range of stoichiometries should exist. Gibbs free energy plot at the temperature of interest would give accurate relative stability of various phases. Assuming room temperature as the temperature of interest, we assess the thermodynamic stability of various phases. Convex hull plot at zero kelvin can give reliable results, as shown in Figure \ref{figHull}. The PV term contribution to the Gibbs free energy is six orders of magnitude lower than internal energy, and the contribution of entropy is also negligible. So we calculated enthalpies of formation per atom for compounds and solid solutions as shown in equation \ref{eqEnthalpyCompounds}  and equation \ref{eqEnthalpySolidSolutions}. The most stable Ti (hcp Ti) and N (N$_2$ gas) phases were considered as reference states in the hull plot.

\begin{equation}
	\Delta H^f\left(Ti_{1-x}N_x\right)=E\left(Ti_{1-x}N_x\right)-\left(1-x\right)\ E\left(Ti\right)-\frac{1}{2}x\ E\left(N_2\right)
	\label{eqEnthalpyCompounds}
\end{equation}

\begin{equation}
	\Delta H^f\left(Ti-xN\right)=E\left(Ti-xN\right)-E\left(Ti\right)-\frac{1}{2}x\ E\left(N_2\right)
		\label{eqEnthalpySolidSolutions}
\end{equation}

\paragraph{}
Where $E(Ti_{1-x}N_x)$ and $E(Ti-xN)$ are the DFT total energies of the considered structures per atom in compounds, and solid solutions, respectively, and $x$ is the atomic fraction of N. $E(Ti)$ and $E(N_2)$ are DFT total energies of Ti and N$_2$. 

\paragraph{}
From the literature, we found the following thermodynamically stable  Ti-N phases: 1) solid solution of N in Ti (around 17-23 at.\% N)  \cite{Xue2014,Dupressoire2021,Bars1983}, it is represented as Ti-xN, where x is the atomic fraction of N, 2) compounds TiN, Ti$_6$N$_5$, Ti$_4$N$_3$, Ti$_3$N$_2$, and Ti$_2$N \cite{Yu2015,Okamoto1993,Xue2014}. Hence, we have considered the following phases in our study: Ti, Ti-0.01N, Ti-0.14N, Ti-0.20N, Ti-0.25N, Ti$_2$N, Ti$_3$N$_2$, Ti$_4$N$_3$, Ti$_6$N$_5$, and TiN. Their detailed structural information like space group, lattice parameters, and Wyckoff positions are given in Table \ref{tableWyckoff}.

\begin{table}[bp]
	\centering
	\caption{The space group, lattice parameters, and Wyckoff positions for various Ti-N phases. In Ti-xN, x is the percentage of N in Ti as an interstitial solid solution, where 0$\le$ x$\le$0.25.}
	\label{tableWyckoff}
	\begin{tabular}{  l       l l l l l l         l l l l l} \hline
	\multirow{2}{*}{Structure} &	\multicolumn{6}{c}{Lattice parameters}     &	\multicolumn{5}{c}{Wyckoff positions} \\
	    	 & a ($\AA$) &	b ($\AA$)	& c ($\AA$)	& $\alpha$ &	$\beta$	& $\gamma$ &	      Atom &	Site&	x	& y &  	z \\ \hline
	    	 
	 TiN (Fm$\bar{3}$m)	&	4.252	&	4.252	&	4.252	&	90	&	90	&	90	&	Ti	&	4a	&	0	&	0	&	0	\\
	 cubic	&		&		&		&		&		&		&	N	&	4b	&	0.5	&	0.5	&	0.5	\\
	 Ti$_6$N$_5$ (C2/m)	&	5.214	&	9.015	&	8.517	&	90	&	144.8	&	90	&	Ti	&	8j	&	0.487	&	0.824	&	0.746	\\
	 monoclinic	&		&		&		&		&		&		&	Ti	&	4i	&	0.001	&	1	&	0.738	\\
	 &		&		&		&		&		&		&	N	&	4h	&	1	&	0.335	&	0.5	\\
	 &		&		&		&		&		&		&	N	&	4g	&	0.5	&	0.166	&	0	\\
	 &		&		&		&		&		&		&	N	&	2c	&	1	&	0	&	0.5	\\
	 Ti$_4$N$_3$ (C2/m)	&	9.934	&	3.018	&	10.302	&	90	&	150.3	&	90	&	Ti	&	4i	&	0.266	&	0.5	&	0.633	\\
	 monoclinic	&		&		&		&		&		&		&	Ti	&	4i	&	0.258	&	0	&	0.885	\\
	 &		&		&		&		&		&		&	N	&	4i	&	0.001	&	0.5	&	0.749	\\
	 &		&		&		&		&		&		&	N	&	2a	&	0	&	1	&	1	\\
	 Ti$_3$N$_2$ (Immm)	&	4.156	&	3.035	&	9.16	&	90	&	90	&	90	&	Ti	&	2d	&	0.5	&	0	&	0.5	\\
	 orthorhombic	&		&		&		&		&		&		&	Ti	&	4j	&	0	&	0.5	&	0.316	\\
	 &		&		&		&		&		&		&	N	&	4i	&	0	&	0	&	0.162	\\
	 Ti$_2$N (P4$_2$/mnm)	&	4.956	&	4.956	&	3.036	&	90	&	90	&	90	&	Ti	&	4f	&	0.203	&	0.797	&	0.5	\\
	 tetragonal	&		&		&		&		&		&		&	N	&	4f	&	0	&	0	&	0	\\
	 Ti-xN (P6$_3$/mmc)	&	2.935	&	2.935	&	4.647	&	90	&	90	&	120	&	Ti	&	2c	&	0.333	&	0.666	&	0.25	\\
	 hexagonal	&		&		&		&		&		&		&	N(x)	&	2a	&	0	&	0	&	0	\\
	 Ti (P6$_3$/mmc)	&	2.935	&	2.935	&	4.647	&	90	&	90	&	120	&	Ti	&	2c	&	0.333	&	0.666	&	0.25	\\
	 hexagonal	&		&		&		&		&		&		&		&		&		&		&		\\  \hline	  
	\end{tabular}
\end{table}

\paragraph{}
To represent a solid solution of N in Ti within the periodic boundary conditions, we took a primitive cell of hcp Ti having two atoms and repeated 4x4x3 along X, Y, and Z directions, forming a supercell with 96 Ti atoms. N is more stable in octahedral voids than tetrahedral voids; the number of octahedral voids is equal to the number of Ti atoms in the supercell. For example, to create Ti-0.14N supercell, we took 96 Ti atoms supercell, and 16 N atoms were placed in octahedral sites such that the average distance between N atoms is maximum. A N atom was placed in a randomly chosen octahedral void in Ti as the starting point,  and the second void was chosen such that it is the farthest from the first void. The third void was chosen such that it is the farthest from the first two voids. The subsequent voids are filled similarly. The same approach is used to construct other supercells representing Ti-xN solid solutions.

\paragraph{}
Figure \ref{figHull} shows the convex hull plot of the Ti-N system. TiN, Ti$_6$N$_5$, Ti$_4$N$_3$, Ti$_3$N$_2$, and Ti$_2$N fall on the convex hull hence are stable, while solid solutions of N in Ti (Ti-xN) are above the convex hull. Although phases above the hull plot are considered unstable, the enthalpies of formation per atom for Ti-0.01N, Ti-0.14N, Ti-0.20N, and Ti-0.25N phases are 0.002 eV, 0.024 eV, 0.033 eV, and 0.043 eV, respectively above the hull. These slight energy differences suggest that they can be stable under suitable conditions. Being part of the graded interface is one such condition that solid solutions could be stable. If the solid solution splits into Ti and Ti$_2$N, this will result in an abrupt change in lattice strain, while the presence of Ti-N solid solutions would provide a gradual change in lattice strain. 	  
	  
\paragraph{}
The reason to obtain a smoothly varying chemically graded interface by controlling N concentration during the growth could be two-fold: 1) phases that are shown in the hull plot have smoothly varying N concentration, and 2) all phases have a close-packed hexagonal arrangement of Ti atoms (Ti$_3$N$_2$, Ti$_4$N$_3$, and Ti$_6$N$_5$ phases are variants of TiN with ordered N vacancies). While forming a graded interface, hexagonal close-packed planes of various phases can epitaxially align.

	  
	  \subsection{Diffusion controlled formation of graded interface}
	  
\paragraph{}	  
If we assume that graded interface can indeed form by just migration of N from TiN to Ti, Gibbs free energy of the Ti/TiN heterostructure should decrease on the migration of N. Assuming N migration occurs in two steps: N vacancy formation in TiN and N interstitial formation in Ti; an overall change in Gibbs free energy, which we refer to as driving force  ($\Delta G^{drive}$) shown in equation \ref{eqDeltaGDrive} is the sum of Gibbs free energy of formation of N vacancies ($\Delta G^{vac}$) given by equation \ref{eqDeltaGVac} and Gibbs free energy of formation of N interstitials ($\Delta G^{int}$) given by equation \ref{eqDeltaGInt}. During the migration process, it is expected that more N vacancies and interstitials are created near the interface; hence, vacancy and interstitial concentration are not uniform. This non-uniformity of N migration makes it very difficult to estimate the driving force. Thus we assume N vacancies are created uniformly on the TiN side, and corresponding N interstitials are created uniformly in Ti. N migration would stop only when the sum of the N vacancy formation energy in Ti$_{1-x}$N$_x$ (Ti in fcc lattice) and interstitial formation energy in Ti-xN (Ti in hcp lattice) becomes positive. We also assume the interface structure would not significantly alter the  driving force ($\Delta G^{drive}$).
	   
	  \begin{equation}
	  	\Delta G^{drive}=\Delta G^{vac}+\Delta G^{int}
	  	\label{eqDeltaGDrive}
	  \end{equation}
	  
	 \begin{equation}
	 	\begin{split}
	 		\Delta G^{vac}=E^{\left[Ti_{1-x}N_x\right]-y}-E^{Ti_{1-x}N_x} +\frac{1}{2}  \mu_{N_2} +P\left(V^{\left[Ti_{1-x}N_x\right]-y}-V^{Ti_{1-x}N_x}\right) \\
	 		-T\left(S_{vib}^{\left[Ti_{1-x}N_x\right]-y}-S_{vib}^{Ti_{1-x}N_x}+S_{con}^{\left[Ti_{1-x}N_{x }\right]-y}-S_{con}^{Ti_{1-x}N_x}\right)
	 	\end{split}
 		\label{eqDeltaGVac}
	 \end{equation}
	  
	 \begin{equation}
	 	\begin{split}
	 		\Delta G^{int}=E^{\left[Ti-xN\right]+y}-E^{Ti-xN}-\frac{1}{2}  \mu_{N_2} + P\left(V^{\left[Ti-xN\right]+y}-V^{Ti-xN}\right) \\
	 		-T\left(S_{vib}^{\left[Ti-xN\right]+y}-S_{vib}^{Ti-xN}+S_{con}^{\left[Ti-xN\right]+y}-S_{con}^{Ti-xN}\right)
	 	\end{split}
 		\label{eqDeltaGInt}
	 \end{equation}
	  
\paragraph{}
Where $x$ is the atomic fraction of N in the Ti-N system, and $y$ is the number of N atoms such that the percentage of N vacancies or interstitials is in the order of ppm. $Ti_{1-x}N_x$ and $Ti-xN$ represent pristine compounds and solid solutions, respectively. $[Ti_{1-x}N_x]-y$ and $[Ti-xN]+y$ represent systems with vacancies and interstitials, respectively. $V$, $S_{vib}$, $S_{con}$ are the volume, vibrational entropy, and configurational entropy, respectively.  $E^{Ti_{1-x}N_x}$ and $E^{Ti-xN}$ are the internal energies of compounds (TiN, Ti$_6$N$_5$, Ti$_4$N$_3$, Ti$_3$N$_2$, and Ti$_2$N) and solid solutions of N in Ti.  $E^{\left[Ti_{1-x}N_x\right]-y}$ and $E^{\left[Ti-xN\right]+y}$ are the internal energies of bulk with N vacancies and N interstitials. $\mu_{N_{2}}$ is the chemical potential of nitrogen, and $P$ is pressure.

\paragraph{}
Adding up equations \ref{eqDeltaGVac} and \ref{eqDeltaGInt}, the term $\frac{1}{2}\mu_{N_2}$  will cancel out. At ambient pressure, the terms, $P\left(V^{\left[Ti_{1-x}N_x\right]-y}-V^{Ti_{1-x}N_x}\right)$ from equation \ref{eqDeltaGVac} and  $P\left(V^{\left[Ti-xN\right]+y}-V^{Ti-xN}\right)$ from equation \ref{eqDeltaGInt}, are insignificant as the difference in the volume of condensed matter due to creation of vacancies or interstitials are negligible. Therefore, we can safely ignore this contribution. Vibrational entropies of $Ti_{1-x}N_x$ with vacancy ($S_{vib}^{\left[Ti_{1-x}N_x\right]-y}$) and without vacancy ($S_{vib}^{Ti_{1-x}N_x}$)  are expected to be similar. Similarly, the difference in vibrational entropies between system $Ti-xN$ with interstitial ($S_{vib}^{\left[Ti-xN\right]+y}$) and without interstitial ($S_{vib}^{Ti-xN}$) should be infinitesimal. Hence, vibrational entropy contribution could be ignored.

\paragraph{}
The configurational entropy, $S_{con}^{Ti_{1-x}N_x}$ is zero as they are ordered compounds. Contribution of configurational entropy to Gibbs free energy $S_{con}^{\left[Ti_{1-x}N_x\right]-y}$ is positive, owing to the creation of vacancy in an ordered compound. $S_{con}^{\left[Ti-xN\right]+y}-S_{con}^{Ti-xN}$ is positive because there are more N interstitials in $[Ti-xN]+y$ than $Ti-xN$. Hence the overall contribution of configurational entropy to Gibbs free energy is negative, thus lowering the $\Delta G^{drive}$. The contribution of configurational entropy is expected to be small, and it is not easy  to accurately account for a range of systems considered, so we ignore its contribution. Thus, the driving force $\Delta G^{drive}$ to form a chemically graded interface can be written as shown in equation \ref{eqDeltaGDriveFinal}.
	  
	  \begin{equation}
	  	\Delta G^{drive}=E^{\left[Ti_{1-x}N_x\right]-y}\ -E^{Ti_{1-x}N_x} +E^{[Ti-xN]+y}-E^{Ti-xN}
	  	\label{eqDeltaGDriveFinal}
	  \end{equation}
  
  \paragraph{}
  To calculate $\Delta G^{drive}$, only the internal energies of defected and pristine structures would suffice and can be reliably calculated using DFT. At 0 K, vacancy formation energy and interstitial formation energy can be calculated according to equations \ref{eqDeltaEVac} and \ref{eqDeltaEInt}, respectively. 
  
  \begin{equation}
  \Delta E^{vac}=E^{\left[Ti_{1-x}N_x\right]-1}-E^{Ti_{1-x}N_x}+\frac{1}{2}E\left(N_2\right)
  \label{eqDeltaEVac}
  \end{equation}

\begin{equation}
	{\Delta E}^{int}=E^{\left[Ti-xN\right]+1}-E^{Ti-xN}-\frac{1}{2}E\left(N_2\right)
	\label{eqDeltaEInt}
\end{equation}

\paragraph{}
Where $E^{\left[Ti_{1-x}N_x\right]-1}$ and $E^{[Ti-xN]+1}$ are the DFT-calculated total energies of the defect supercell containing one N vacancy and one N interstitial, respectively. $E^{Ti_{1-x}N_x}$ and $E^{Ti-xN}$ are the DFT-calculated total energies of the pristine supercells with compounds and solid solutions. $E\left(N_2\right)$ is the DFT total energy of N$_2$ molecule. The sum of vacancy formation energy and interstitial formation energy is equal to $\Delta G^{drive}$.

\paragraph{}
Vacancy and interstitial formation energies are defined for a small vacancy or interstitial concentration in a few ppm. Thus supercell containing a few million atoms is required. Since it is not practical to model millions of atoms using density functional theory, we have modeled at least 1.04\% N vacancies or interstitials and confirmed that further reducing the percentage of vacancies and interstitials does not significantly affect their formation energies. Even if the concentration of defects is reduced by half, formation energies change by a maximum of 100 meV. We would also like to point out that supercells considered are large, and the images of vacancies or interstitials are at least 10 $\AA$ apart.

\paragraph{}
In principle, one should calculate vacancy formation energy in all Ti$_{1-x}$N$_x$ and interstitial formation energy in all Ti-xN. Given computational limitation, we only calculate for a few stable compounds identified from hull plot: TiN, Ti$_6$N$_5$, Ti$_4$N$_3$, and Ti$_3$N$_2$. We correspondingly chose Ti, Ti-0.14N, Ti-0.20N, and Ti-0.25N such that the fraction of N vacant/filled sites in TiN/Ti, Ti$_6$N$_5$/Ti-0.14N, Ti$_4$N$_3$/Ti-0.20N, Ti$_3$N$_2$/Ti-0.25N are the same. The fraction of N vacant sites is the ratio of the unfilled N sites to the available N sites and the fraction of N-filled sites is the ratio of the filled N sites to the available N sites. All systems were modeled using supercells containing 96 Ti atoms; thus, 96 octahedral sites are available for N to occupy.

\begin{table}[tp]
	\caption{Octahedral N vacant and filled sites in the Ti-N systems.}
	\label{tableSiteFraction}
	\centering
	\begin{tabular}{  l l l l l l l        l l l l l l l   }  \hline 
		\multirow{2}{*}{Compound}  & \multicolumn{2}{c}{Pristine} & \multicolumn{2}{c}{Defect} & Vacant & \% Vacant &
		Solid  & \multicolumn{2}{c}{Pristine} & \multicolumn{2}{c}{Defect} & Filled & \% Filled
		\\ 
		& Ti & N & Ti & N &   N sites    & N sites  & 	        Solution  & Ti & N & Ti & N & N sites & N sites \\ \hline
		
		TiN	&	96	&	96	&	96	&	95	&	1	&	1.04	&	Ti	&	96	&	0	&	96	&	1	&	1	&	1.04	\\
		Ti$_6$N$_5 $	&	96	&	80	&	96	&	79	&	17	&	17.71	&	Ti-0.14N&	96	&	16	&	96	&	17	&	17	&	17.71	\\
		Ti$_4$N$_3$	&	96	&	72	&	96	&	71	&	25	&	26.04	&	Ti-0.20N&	96	&	24	&	96	&	25	&	25	&	26.04	\\
		Ti$_3$N$_2$	&	96	&	64	&	96	&	63	&	33	&	34.38	&	Ti-0.25N&	96	&	32	&	96	&	33	&	33	&	34.38	\\ \hline		 												  
	\end{tabular}
\end{table}

\begin{figure}[tp]
	\centering
	\includegraphics[width=10cm]{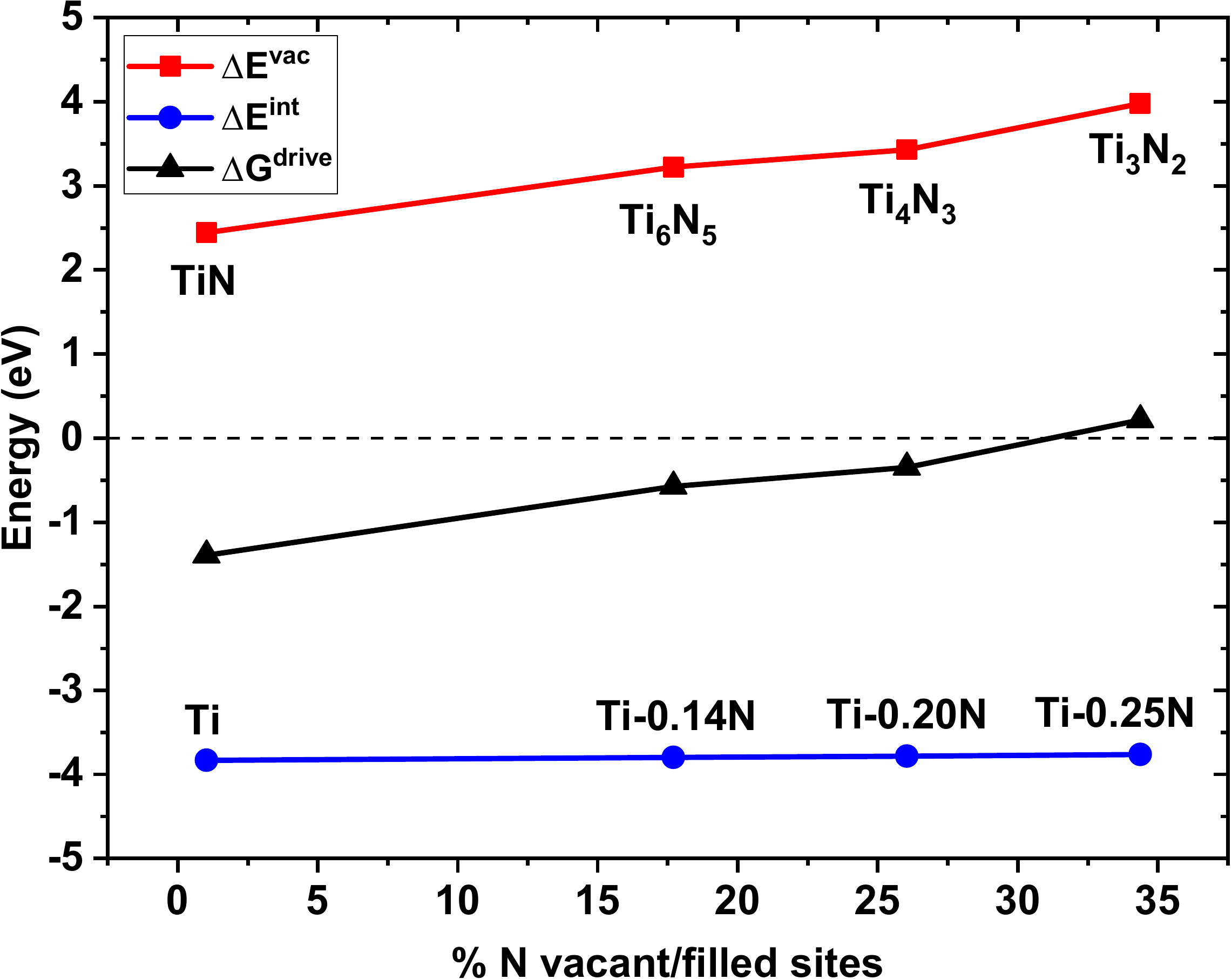}
	\caption{Vacancy/interstitial formation energy as a function of atomic \% N vacant/filled sites. Red squares represent vacancy formation energies ($\Delta E^{vac}$), and blue circles represent interstitial formation energies ($\Delta E^{int}$) of Ti-xN phases, where x is the concentration of N. The sum of vacancy and interstitial formation energies is the driving force ($\Delta G^{drive}$) as indicated in black upper triangles.}
	\label{figDrive}
\end{figure}

\paragraph{}
Table \ref{tableSiteFraction} shows the number of Ti and N atoms in the pristine and defect supercells. Defect supercells were constructed by removing/adding one N atom from pristine supercells. The vacancy formation energy of Ti-N compounds could depend on the choice of Wyckoff positions for removal of N. So we removed N from all available Wyckoff positions, and vacancy formation energy shown in figure \ref{figDrive} corresponds to the lowest formation energy because the first N to come out will be the one which is easiest to remove. In Ti$_6$N$_5$, N occupies 2c, 4g, and 4h Wyckoff positions, and the N removed from the 4g position has the lowest energy value of 3.22 eV (2c has 3.25 eV and 4h has 3.29 eV). Similarly, in Ti$_4$N$_3$, N occupies 2a and 4i Wyckoff positions, and the N removed from the 2a position has the lowest energy value of 3.43 eV (4i has 3.65 eV).

\begin{figure}[tp]
	\centering
	\includegraphics[width=\textwidth]{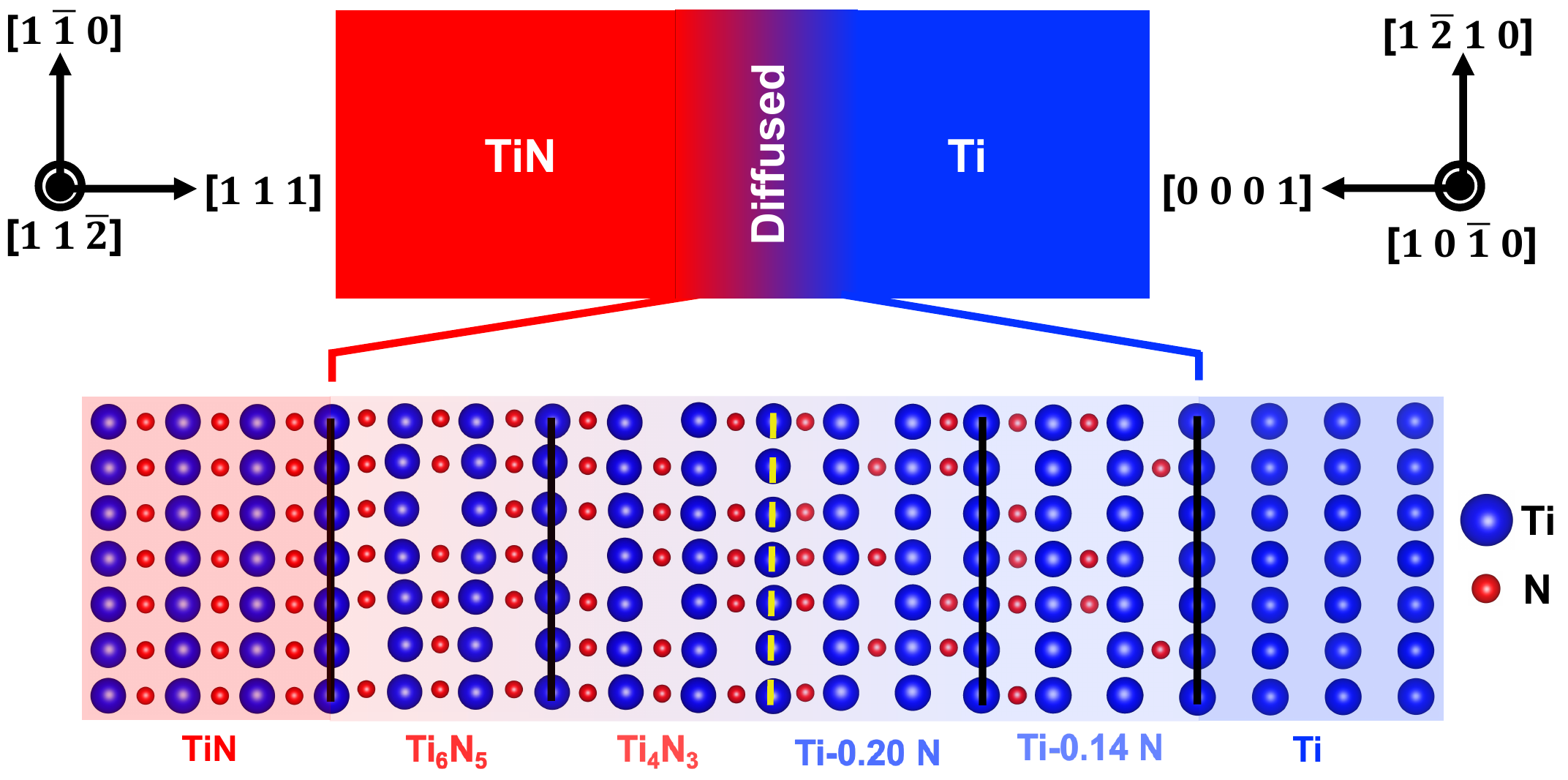}
	\caption{Schematic representation of diffused interface and atomic representation of the diffused Ti/TiN interface composing of Ti-N phases considered in the calculation of driving force. Ti (0001) plane is parallel to TiN (111) plane, and both are parallel to close-packed planes of all other phases. The average Ti-Ti bond length in the close-packed plane of all phases is assumed to be the same in the atomic representation.}
	\label{figDiffused}
\end{figure}

\paragraph{}
Figure  \ref{figDrive} shows vacancy and interstitial formation energies and driving force. Vacancy formation energy increases as more and more N is removed from FCC Ti. In contrast, interstitial formation energy does not change significantly as N is added to HCP Ti, at least for the range considered.  The driving force for N diffusion from FCC Ti to HCP Ti is most negative (-1.39 eV/atom) when 1.04\% of N atoms in TiN is removed, and the corresponding percentage Ti octahedral sites are filled by N. Ti$_6$N$_5$ and Ti-0.14N correspond to 17.71\% N vacant and filled sites in FCC and HCP Ti, respectively. The driving force for N atom to diffuse from FCC lattice to HCP lattice now reduces to -0.58 eV/atom. As more N diffuses to HCP Ti, the driving force further reduces to -0.35 eV/atom and becomes positive (0.21 eV/atom) for 26.04\% and 34.38\% N atoms changing sides, respectively. As the diffusion of N would be thermodynamically unfavorable when the driving force becomes positive, it can be inferred from Figure \ref{figDrive} that a maximum of 23 at.\% of N can migrate. 
\paragraph{}
Our calculations suggest that starting from a sharp interface by creating N vacancies in TiN and their diffusion into Ti is thermodynamically favorable, and the driving force decreases as more N diffuse. A sufficient amount of N vacancies in TiN near the interface transforms the region to Ti$_6$N$_5$ and then progressively to Ti$_4$N$_3$; on the other side, more and more N interstitials lead to the formation of Ti-xN (Ti-0.20N and Ti-0.14N) phases. Although our results suggest that migration should happen until all TiN transforms to the Ti$_4$N$_3$ phase, the kinetics of migration of N would hinder in completion of the transformation, and such transformation is expected to be only near the interface. This should result in: 1) on TiN side phases with the most amount of N vacancies would be closest to the interface like Ti$_4$N$_3$, followed by Ti$_6$N$_5$ and TiN; 2) on Ti side N rich solid solutions would be closest to the interface and N concentration would decrease away from the interface. The interplay of driving force and kinetics thus leads to forming a chemically graded interface, as shown in Figure \ref{figDiffused}. We want to point out that we only discuss phases that we calculated to estimate driving force, but we believe there will be other phases that have N concentration between the phases examined. They can form by forming vacancies in ordered structures like TiN, Ti$_6$N$_5$, and Ti$_4$N$_3$ or varying N interstitial concentrations in solid solutions. 

\paragraph{}
As Ti$_4$N$_3$ and Ti$_6$N$_5$ have rocksalt structures with ordered N vacancy, the proposed diffused interface can occur by diffusion of N from TiN side to Ti, without changing the atomic position of Ti. As the diffusion of N controls the formation of a diffused interface, temperature and exposure time will dictate the width of the diffused interface. Ti generally grows along $<$0001$>$ direction and TiN grows along $<$111$>$ direction, thus ${\left(111\right)_{TiN}\ ||\ \left(0001\right)}_{Ti}$. Since the atomic arrangement of Ti does not change on the formation of the diffused interface, close-packed planes of other phases will also be parallel to $\left(111\right)_{TiN}$ and $\left(0001\right)_{Ti}$.

		  	\subsubsection{Misfit among Ti-N phases in 3-D Ti/TiN interface}

\begin{figure}[bt]
	\centering
	\includegraphics[width=\textwidth]{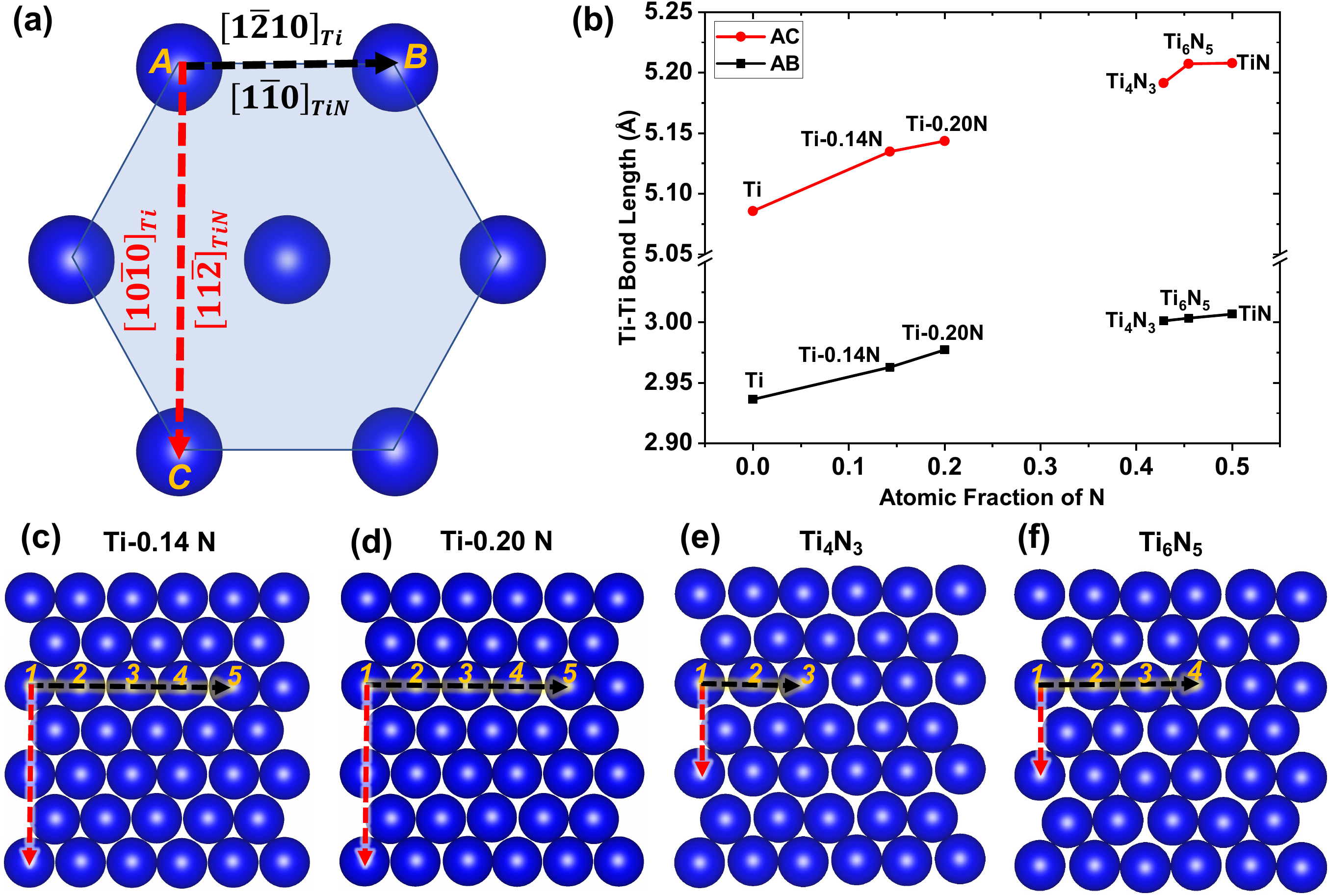}
	\caption{Atomic structure of close-packed (0001) plane of HCP Ti and (111) plane of rocksalt TiN. Blue spheres represent Ti atoms. AB and AC are the Ti-Ti bond lengths along $\left[10\bar{1}0\right]_{Ti}/\left[11\bar{2}\right]_{TiN}$ and $\left[1\bar{2}10\right]_{Ti}/\left[1\bar{1}0\right]_{TiN}$ directions, respectively. (b) Variation of Ti-Ti bond lengths along AB and AC directions. (c)-(f) represents the atomic structure of close-packed planes and shows two orthonormal lattice vectors used to calculate average AB and AC distances.}
	\label{figInPlaneBondLength}
\end{figure}

\paragraph{}
In a sharp Ti/TiN interface, misfit is accommodated in one plane near the interface if not at the interface by forming misfit dislocations. The atomistic structure of close-packed planes of hcp Ti and rocksalt TiN is shown in Figure \ref{figInPlaneBondLength}a. In close-packed planes of Ti and TiN, $\left[1\bar{1}0\right]_{TiN}\ ||\ \left[11\bar{2}0\right]_{Ti}$ and $\left[11\bar{2}\right]_{TiN}\ ||\ \left[10\bar{1}0\right]_{Ti}$. AB and AC are Ti-Ti bond lengths along $\left[1\bar{1}0\right]_{TiN}/\left[11\bar{2}0\right]_{Ti}$ and $\left[11\bar{2}\right]_{TiN}/\left[10\bar{1}0\right]_{Ti}$, respectively. The calculated percentage difference in the Ti-Ti bond length in Ti and TiN along the two orthonormal directions is 2.34\%, and percentage difference along these directions has been used to assess misfit strain across the Ti/TiN interface \cite{Yang2017}. 
		  	
\paragraph{}
All the phases mentioned in Figure  \ref{figDiffused} are expected to be a part of the chemically graded interface with the hexagonal arrangement (not necessarily regular hexagon) of Ti atoms, so equivalent Ti-Ti bond lengths AB and AC can be calculated for all phases. These lengths can serve as good indicators of misfit strain among various phases present in the Ti/TiN graded interface. The phases other than Ti and TiN do not have a regular hexagonal arrangement (as the Ti-Ti bond length depends on the distribution of N atoms around the chosen bond). The atomic structure of the close-packed planes of Ti-0.14N, Ti-0.20N, Ti$_4$N$_3$, and Ti$_6$N$_5$ are shown in Figure \ref{figInPlaneBondLength} (c)-(f). We identified orthonormal lattice vectors in the close-packed plane along the directions similar to $\left[1\bar{1}0\right]_{TiN}/\left[11\bar{2}0\right]_{Ti}$ and $\left[11\bar{2}\right]_{TiN}/\left[10\bar{1}0\right]_{Ti}$. These lattice vectors were used to calculate average AB and AC distances.

\paragraph{}
The measured AB and AC lengths for various phases are plotted in Figure \ref{figInPlaneBondLength}b. There is a gradual increase in length of both AB and AC as N concentration increases. This increase points to the possibility of misfit dislocations being distributed over several planes, between various phases. Thus dislocation density in any given plane would be significantly lower than the sharp interface plane of Ti/TiN, which could result in unusual interaction of dislocations and cracks with the interface.

	  		\subsubsection{Mechanical Properties of Ti-N phases at Ti/TiN interface}

\paragraph{}	  	
Heterostructures of Ti/TiN, in the form of a single-layer TiN coating on Ti alloys or Ti/TiN nano-multilayer coating, have been extensively explored for various structural applications, like wear \cite{Cheng2010} and erosion resistance \cite{Cao2019}. Mechanical properties of the heterostructure would depend on the mechanical properties of the individual phases in the heterostructure and its variation across the interface. We calculate mechanical properties such as bulk modulus (B), shear modulus (G), Young’s modulus (E), Poisson’s ratio ($\nu$), Pugh’s (G/B) ratio, and Vicker’s hardness (H$_V$). The calculated elastic constants C$_{ij}$ are reported in Table \ref{tableElasticConst}. The second-order elastic constants, C$_{ij}$, were calculated using the energy-strain method implemented in vaspkit \cite{Wang2021}  code by straining the lattice from -1.5\% to +1.5\% with a 0.5\% step size for each independent elastic constant. Voigt-Reuss-Hill (VRH) approximation \cite{Voigt1928,Reuss1929,Hill1952} was employed to obtain bulk modulus, shear modulus, Young’s modulus, and Poisson’s ratio. The theoretical Vickers hardness Hv was estimated by using Chen’s model \cite{Chen2011}, according to equation \ref{eqVickersHardness}.

\begin{equation}
	H_V=2\left(\kappa^2G\right)^{0.585}-3
	\label{eqVickersHardness}
\end{equation}	  	
Where $\kappa$ is the Pugh’s ratio,  $\kappa$= G/B.
	  	
\begin{figure}[tp]
	\centering
	\includegraphics[width=\textwidth]{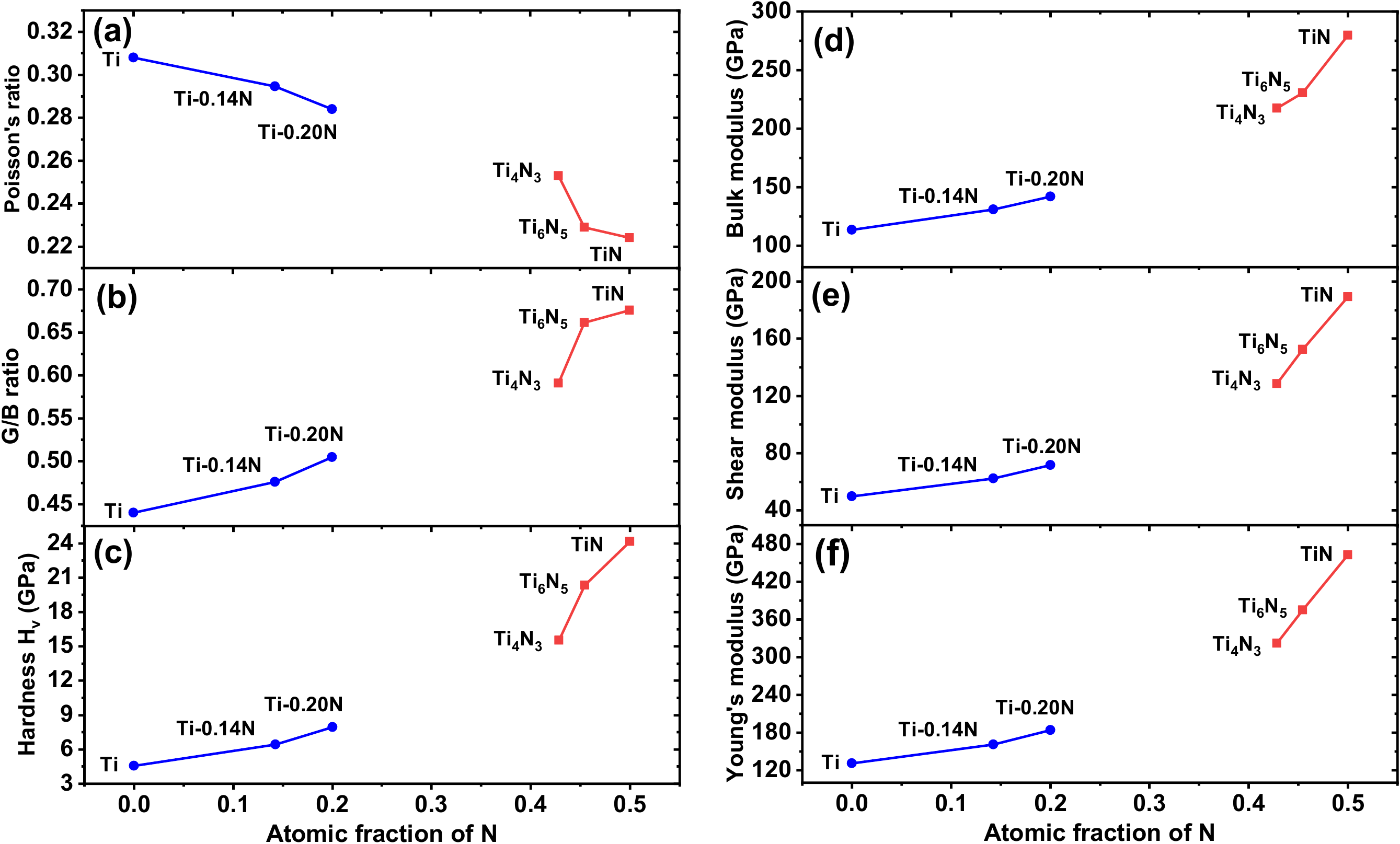}
	\caption{Variation of Poisson’s ratio, G/B ratio, Vickers hardness Hv, bulk modulus B, shear modulus G, and Young’s modulus E, of Ti-N phases expected to be present in chemically graded interface, as a function of N concentration.}
	\label{figMechProp}
\end{figure}	  	
	  	
\paragraph{}
Poisson’s ratio decreased from Ti to TiN, indicating the material becoming more covalent as the concentration of N increases in Ti. The value of  $\sim$0.2 and $\sim$0.4 indicates strong directional covalent and metallic bonding, respectively \cite{Yu2015,Pugh1954}. A value of Pugh’s (G/B) ratio of $>$0.57 indicates brittleness suggesting Ti$_4$N$_3$, Ti$_6$N$_5$, and TiN are brittle and solid solutions of N in Ti, including Ti, are ductile. In line with Poisson’s and Pugh’s ratios, we observed a gradual increase in Vickers hardness from Ti to TiN.

\paragraph{}
Figure \ref{figMechProp} (d)-(f) shows a gradual variation of bulk modulus, shear modulus, and Young’s modulus as a function of N concentration of phases in the graded interface. All of them increase smoothly as the concentration of N increases. It has been shown that smooth variation of modulus at micrometer length scale across the interface improves heterostructure properties like wear resistance \cite{Suresh2001} and crack suppression \cite{Jitcharoen1998}. Hence a similar gradation at atomic scale could improve properties of heterostructures containing the graded interface. In fact, we suspect that some of the properties of Ti/TiN heterostructures could be due to the 3-D interface. 

\begin{table}[btp]
	\caption{The calculated elastic constants $C_{ij}$ (GPa) of the Ti–N phases.}
	\label{tableElasticConst}
	\centering
	\begin{tabular}{ l l l l l l l l l l l l l l }  \hline
		Phase	&	C$_{11}$	&	C$_{12}$	&	C$_{13}$	&	C$_{15}$	&	C$_{22}$	&	C$_{23}$	&	C$_{25}$	&	C$_{33}$	&	C$_{35}$	&	C$_{44}$	&	C$_{46}$	&	C$_{55}$	&	C$_{66}$ \\ \hline
		Ti	&	182	&	81	&	76	&		&		&		&		&	191	&		&	45	&		&		&	\\
		Ti-0.14N	&	204	&	101	&	84	&		&		&		&		&	230	&		&	70	&		&		&	\\
		Ti-0.20N	&	230	&	105	&	86	&		&		&		&		&	262	&		&	75	&		&		&	\\
		Ti$_4$N$_3$	&	370	&	131	&	149	&	-17	&	400	&	112	&	1	&	410	&	-23	&	132	&	7	&	116	&	137 \\
		Ti$_6$N$_5$	&	423	&	131	&	135	&	-16	&	425	&	134	&	15	&	428	&	2	&	156	&	15	&	169	&	150 \\
		TiN	&	592	&	123	&		&		&		&		&		&		&		&	164	&		&		&	\\ \hline
	\end{tabular}
\end{table}

\section{Conclusions}
Using first-principles DFT calculations, we show that the atomically chemically graded interface (which we refer to as 3-D interface) between Ti/TiN is thermodynamically stable over a sharp interface. To establish the existence of a 3-D interface, we define the term driving-force, which is the sum of N vacancy formation energy in TiN and N interstitial formation energy in Ti. The concept of driving-force can be used to assess other metal/ceramic interfaces for the possibility of forming an atomically chemically graded interface. We show that diffusion of N from TiN to Ti side alone is required to form a 3-D interface. We also show that a range of stoichiometry of the Ti-N system is stable; this explains the experimental effort to make a chemically graded interface by controlling N partial pressure during the growth of Ti/TiN to form a graded interface. 

We show that there is smooth variation in lattice parameter and mechanical properties like (bulk modulus, shear modulus, Young’s modulus, and hardness) across the interface of Ti/TiN. Such structure and properties can result in novel mechanical properties of Ti/TiN heterostructures. Gradation of structure and properties in heterostructures at micrometer length scale has been shown to improve their properties; a similar improvement could be expected in heterostructures having such graded interface. Our findings suggest that several promising properties of Ti/TiN heterostructure reported in literature could be due to a chemically graded interface. Any attempt to understand Ti/TiN heterostructure should involve an atomically chemically graded interface, otherwise assumed to be atomically sharp.

\subsection*{AUTHOR INFORMATION}
\subsection*{Corresponding Author}
*Author to whom correspondence should be addressed.  Electronic mail:  satyesh@iitm.ac.in

\subsection*{Author Contributions}
The manuscript was written through contributions of all authors. All authors have given approval to the final version of the manuscript.
  
\subsection*{Funding}

This work was supported by Science and Engineering Research Board (SERB), New Delhi (Grant number: SRG/2019/000455) and Ministry of Human Resource and Development (Grant number: SB20210844MMMHRD008277)
  
\subsection*{Notes}
The authors declare no competing financial interest.

\begin{acknowledgement}
	We acknowledge the use of the computing resources at High Performance Computing Environment (HPCE), IIT Madras. 
\end{acknowledgement}

\bibliography{bibliography.bib} 

\end{document}